\documentclass[aps,prd,twocolumns,eqsecnum,preprintnumbers,showpacs,amsmath,amssymb]
{revtex4}

\usepackage{graphicx}% Include figure files
\usepackage{dcolumn}% Align table columns on decimal point
\usepackage{bm}% bold math

\begin{document}

\title{II. THE MASS GAP AND SOLUTION OF THE \\ QUARK CONFINEMENT PROBLEM IN QCD}

\author{V. Gogokhia}
\email[]{gogohia@rmki.kfki.hu}

\affiliation{HAS, CRIP, RMKI, Depart. Theor. Phys., Budapest 114,
P.O.B. 49, H-1525, Hungary}

\date{\today}
\begin{abstract}

We have investigated a closed system of equations for the quark
propagator, obtained earlier within our general approach to QCD at
low energies. It implies quark confinement (the quark propagator
has no pole, indeed), as well as the dynamical breakdown of chiral
symmetry (a chiral symmetry preserving solution is forbidden).
This system can be solved exactly in the chiral limit. We have
established the space of the smooth test functions (consisting of the Green's
functions for the quark propagator and the corresponding quark-gluon
vertex) in which our generalized function (the confining gluon
propagator) becomes a continuous linear functional. It is a linear
topological space $K(c)$ of the infinitely differentiable
functions (with respect to the dimensionless momentum variable $x$),
having compact support in the region $x \leq c$. We
develop an analytical formalism, the so-called chiral perturbation
theory at the fundamental quark level, which allows one to find
explicit solution for the quark propagator in powers of the light
quark masses. We also develop an analytical formalism, which
allows one to find the solution for the quark propagator in the
inverse powers of the heavy quark masses. It justifies the use for
the heavy quark propagator its free counterpart up to terms of the
order $1/m^3_Q$, where $m_Q$ is the heavy quark mass. So this
solution automatically possesses the heavy quark spin-flavor
symmetry.
\end{abstract}

\pacs{ 11.15.Tk, 12.38.Lg}

\keywords{}

\maketitle

\section{Introduction}

In our previous works \cite{1,2} we have formulated the
intrinsically nonperturbative (INP) QCD as the true theory of QCD
at low energies. In general, it is defined as QCD from which
all types and at all level the perturbative (PT) contributions
("PT contaminations") are to be subtracted. This theory makes it
possible to calculate the physical observables/processes in
low-energy QCD from first principles in a self-consistent way. One
of the main roles in the realization of this program belongs to
the solution for the gluon Green's function which describes their
propagation in the QCD vacuum. In the presence of a mass gap
responsible for the true NP QCD dynamics it has been exactly
established \cite{1}. In the subsequent paper \cite{2} we apply
this solution to the quark, ghost and ghost-quark sectors in order
to derive the system of equations for the quark propagator, namely

\begin{eqnarray}
S^{-1} (p)&=& S_0^{-1} (p)+ \Lambda_{NP}^2 \Gamma_\mu(p,0) S(p)
\gamma_\mu,
\nonumber\\
\Gamma_\mu(p,0) &=& id_\mu S^{-1}(p) - S(p) \Gamma_\mu(p,0)
S^{-1}(p).
\end{eqnarray}
The obtained system of equations (1.1) is exact, i.e., no
approximations/truncations have been made so far. Formally it is
valid in the whole energy/momentum range, but depends only on the
mass gap $\Lambda_{NP}^2$ responsible for the true NP QCD
dynamics. It is free of all types of the PT contributions
("PT contaminations") at the fundamental quark-gluon-ghost level.
As mentioned in Ref. \cite{2}, the free PT quark propagator is not
to be subtracted in order to maintain the chiral limit physics in
QCD. This limit is important to correctly understand the structure of
QCD at low energies. Also, it is manifestly gauge-invariant, i.e.,
does not depend explicitly on the gauge-fixing parameter.

The main purpose of this work is to solve this system explicitly
and show that it describes the confining quark propagator, indeed.
We will show that it possesses the property of the dynamical
breakdown of chiral symmetry as well. Concluding, let us only note
that a very preliminary derivation of the system of equations (1.1)
has been made in our earlier papers \cite{3,4}.

\section{Solution}

The simplest way to solve the ground system of equations (1.1) is
to represent the corresponding proper quark-gluon vertex at zero
momentum transfer as its decomposition into the four independent
matrix structures, i.e.,

\begin{equation}
\Gamma_\mu(p,0) = \gamma_\mu F_1(p^2)+ p_\mu F_2(p^2)
 + p_\mu \hat p F_3(p^2)   + \hat p\gamma_\mu F_4(p^2).
\end{equation}
The Euclidean version of the chosen parametrization for the full
quark propagators is as follows:

\begin{equation}
iS(p)= \hat p A(p^2) - B (p^2),
\end{equation}
so its inverse is

\begin{equation}
iS^{-1}(p)= \hat p \overline A(p^2) +  \overline B (p^2),
\end{equation}
where

\begin{eqnarray}
\overline A(p^2) &=& A(p^2) E^{-1}(p^2), \nonumber\\
\overline B(p^2) &=& B(p^2) E^{-1}(p^2), \nonumber\\
E(p^2) &=& p^2 A^2(p^2) + B^2(p^2).
\end{eqnarray}

Substituting the representation (2.1) into the second equation of
the system (1.1), on account of the previous relations, and doing
some tedious algebra of the $\gamma$-matrices in the 4D Euclidean
space we can express the scalar functions $F_i(p^2) \ (i = 1,2,3,4
)$ in terms of the quark propagator form factors $A(p^2)$ and
$B(p^2)$ as follows:

\begin{eqnarray}
 F_1(p^2) &=& - {1 \over 2} \overline A(p^2), \nonumber \\
 F_2(p^2) &=& -  \overline B'(p^2) + F_4(p^2), \nonumber \\
 F_3(p^2) &=& - \overline A'(p^2) , \nonumber \\
 F_4(p^2) &=&  - {1 \over 2} \overline A^2(p^2) \overline B^{-1}(p^2).
\end{eqnarray}
 Here the prime denotes the derivative with respect to the
 Euclidean momentum variable  $p^2$.

Substituting the solution  for the ST identity (2.5) back into the
quark SD equation, which is the first of the equations in the ground
system (1.1), and again doing the above-mentioned rather tedious
but well-known algebra and introducing further the dimensionless
variables and functions

\begin{equation}
A(p^2) =  \Lambda_{NP}^{-2} A(x), \qquad B(p^2) =
\Lambda_{NP}^{-1} B(x),\qquad
 x = p^2/{\Lambda_{NP}^2},
\end{equation}
one finally obtains that the ground system of equations (1.1)
is reduced to

\begin{eqnarray}
 x A' &=& - (2 + x) A - 1 - \bar m_0 B, \nonumber\\
2B B' &=& -3 A^2  - 2 (B - \bar m_0 A)B.
\end{eqnarray}
Here and everywhere below $A \equiv A(x)$, $B \equiv B(x)$, and
the prime denotes the derivative with respect to the Euclidean
dimensionless momentum variable $x$. We also introduce the
notation for the dimensionless current quark mass as follows:
$\bar m_0 = m_0 / \Lambda_{NP}$, and omitting flavor index, for
simplicity. As mentioned above, this system and consequently the
initial system (1.1) for the first time has been obtained in Refs.
\cite{3,4}.

 The formal exact solution of the ground system (2.7) for the dynamically
 generated quark mass function $B(x)$ is

\begin{equation}
 B^2(c, \bar m_0; x) =  \exp(- 2x)
\int^{c}_x \exp(2x') \tilde{\nu}(x')\, dx' ,
\end{equation}
where $c$ is the constant of integration. Not losing generality,
it can be fixed as $c= p^2_c / \Lambda^2_{NP}$, where $p^2_c$ is
some constant momentum squared. Also

\begin{equation}
\tilde{\nu}(x) = A(x) [3 A(x) +2 \nu(x)]
\end{equation}
with

\begin{equation}
\nu (x) = xA'(x) + (2 + x)A(x) + 1 = - \bar m_0 B(x).
\end{equation}
Then the equation for determining the $A(x)$ function becomes

\begin{equation}
{d\nu^2 (x) \over dx}+ 2 \nu^2(x)= - \tilde{\nu}(x)\bar m^2_0.
\end{equation}
It is interesting to note that the last equation can be exactly solved up to
the terms of the order $\bar m_0$, and may be even up to the terms of the order
$\bar m^2_0$, since the function $\nu (x)$ is of the order
$\bar m_0$ itself (see Eq. (2.10)).

\subsection{Chiral limit}

In the chiral limit ($\bar m_0 = 0$) the ground system (2.7) is
reduced to

\begin{eqnarray}
 x A'_0 &=& - (2 + x) A_0 - 1, \nonumber\\
2B_0 B'_0 &=& -3 A^2_0  - 2 B^2_0,
\end{eqnarray}
which can be solved exactly. The exact solution for the $A_0(x)$
function is

\begin{equation}
A_0(x) =  {1 \over x^2} \left\{ 1 - x - \exp(-x) \right\},
\end{equation}
while for the dynamically generated quark mass function $B_0(x)$
the exact solution is

\begin{equation}
 B^2_0(c_0; x) = 3 \exp(- 2x)
\int^{c_0}_x \exp(2x') A^2_0(x')\, dx' ,
\end{equation}
where $c_0 = p_0^2 / \Lambda_{NP}^2$ is the constant of
integration, and $p^2_0$ is some constant momentum squared for the
chiral limit case. The solution (2.13) has the correct asymptotic
properties. It is regular at small $x$ and asymptotically
approaches the free propagator at infinity. This behavior can formally be
achieved in two ways: $p^2 \rightarrow \infty$ at fixed
$\Lambda^2_{NP}$ and/or by $\Lambda^2_{NP} \rightarrow 0$ as well.
Let us remind that the last limit is known as the PT one. It is
interesting to note that the chiral limit in terms of the
dimensionless mass scale parameter $\bar m_0=0$ can be only
achieved in one way: $m_0 \rightarrow 0$ at fixed $\Lambda_{NP}$,
since at $m_0$ fixed the mass gap $\Lambda_{NP}$ cannot go to
infinity (it is either finite or zero, by definition \cite{1,2}).

Both solutions (2.13) and (2.14) are not, in principle, entire
functions. The functions $A_0(x)$ and $B_0(c_0; x)$ have removable
singularities at zero, i.e., they are finite at zero points. In
addition, the dynamically generated quark mass function $B_0(c_0;
x)$ also has the algebraic branch points at $x=c_0$ and at
infinity (at fixed $c_0$). As in the general (non-chiral) case for
the quark mass function (2.8) at $x=c$, these unphysical
singularities are caused by the inevitable ghost contributions in
the covariant gauges.

As was mentioned above, $A_0(x)$ automatically has a correct
behavior at infinity (it does not explicitly depend on the
constant of integration, since it was specified in order to get a
regular at zero solution). In the PT limit ($\Lambda^2_{NP}
\rightarrow 0$) the constants of integration $c_0, c$ and variable
$x$ go to infinity uniformly ($c_0, c, x \rightarrow \infty$), so
the dynamically generated quark mass functions (2.8) and (2.14)
identically vanish in this limit. Obviously, we have to keep the
constants of integration $c_0, c$ arbitrary but finite in order to
obtain a regular at zero point solutions. The problem is that if
$c_0, c = \infty$, then the corresponding solutions do not exist
at all at any finite $x$, in particular at $x=0$.

Concluding, let us note that the singular at zero point exact
solutions of the system (2.12) also exist. It is easy to check
that the exact solution $A_0(x) = (1 /x^2)(1 - x)$ automatically
satisfies it. Substituting it into the Eq. (2.14), one obtains the
exact solution for the dynamically generated quark mass function,
which will be singular at zero point as well. However, similar to
the ghost self-energy \cite{2}, the singular at zero solution for
the quark propagator should be excluded from the consideration,
since the smoothness properties of the test functions will be
compromised in this case.

\section{Quark confinement}

In principle, it is possible to develop the calculation schemes in
different modifications, which will give the solution of the
ground system (2.7) step by step in powers of the light current
quark masses as well as in the inverse powers of the heavy quark
messes (see sections below). The important observation, however,
is that the formal exact solution (2.8) exhibits the algebraic
branch point at $x=c$, which completely $excludes \ a \ pole-type
\ singularity$ at any finite point on the real axis in the
$x$-complex plane whatever solution for the $A(x)$ function might
be. Thus the solution for the quark propagator cannot be presented
as an expression having finally a pole-type singularity at any
finite point $p^2 = - m^2$ (Euclidean signature), i.e.,

\begin{equation}
S(p) \neq {const \over \hat p + m},
\end{equation}
certainly satisfying thereby the first necessary condition of
quark confinement formulated at the fundamental quark level as the
absence of a pole-type singularity in the quark propagator
\cite{1}. A quark propagator may or may not be an entire function,
but in any case a pole-type singularity has to disappear. This is
a general feature of quark confinement, which holds in any gauge
(see our paper \cite{1} and references therein).

 In order to confirm this, let us assume the opposite to
Eq. (3.1), i.e., that the quark propagator within our approach
may have a pole-type singularity like the electron propagator has
in quantum electrodynamics (QED) (see Eq. (3.4) below). In terms
of the dimensionless quark form factors, defined in Eq. (2.6),
this means that in the neighborhood of the assumed pole at $x =
-m^2$ (Euclidean signature), they can be presented as follows:

\begin{eqnarray}
A(x) &=& { 1 \over (x + m^2)^{\alpha}} \tilde{A} (x), \nonumber\\
B(x) &=& { 1 \over (x + m^2)^{\beta}} \tilde{B} (x),
\end{eqnarray}
where $\tilde{A}(x)$ and $\tilde{B}(x)$ are regular at a pole,
while ${\alpha}$ and ${\beta}$ are, in general, arbitrary with $Re
{\alpha},{\beta} \geq 0$. However, substituting these expansions
into the system (2.7) and analyzing it in the neighborhood of the
assumed pole, one can immediately conclude in that the
self-consistent system for the quantities with tilde exists if and
only if
\begin{equation}
\alpha = \beta = 0.
\end{equation}
In other words, our system (2.7) does not admit a pole-type
singularities in the quark propagator in complete agreement with
the above-mentioned.

This point deserves a more detail discussion, indeed. The infrared
(IR) asymptotic of the electron propagator in QED is \cite{5}
(Minkowski signature)

\begin{equation}
S(p) \sim {1 \over (p^2 - m^2)^{1 +\beta}},
\end{equation}
where $\beta = \alpha(\xi -3) / 2 \pi$ and here $\alpha$ is the
renormalized charge and $\xi$ is the gauge-fixing parameter. Thus
instead of a simple pole, it has a cut whose strength can be
varied by changing $\xi$. However, there is, in general, a
pole-type singularity at the electron mass $m$, indeed, i.e., in
QED there is no possibility to escape a pole-type singularity in
the electron Green's function. Contrary to QED, our general
solution (2.8) has no pole-type singularities, only the branch
point at $x=c$, and the constant of integration $c$ may, in
general, depend implicitly on $\xi$. At the same time, it is
obvious that the existence of a branch point itself does not
depend explicitly on the gauge choice. Thus the absence of a
pole-type singularity in QCD in the same way is gauge-invariant
as the existence of a pole-type singularity at the electron mass
in QED. This may be used indeed to differentiate QCD from QED and
vice versa. The gauge invariance of the above-mentioned first
necessary condition of quark confinement should be precisely
understood in this sense.

The second sufficient condition formulated at the hadronic level
as the existence of a discrete spectrum only (no continuum in the
spectrum) \cite{6} in the bound-state problems within the
corresponding BS formalism  is obviously beyond the scope of the
present investigation. Let us only note here, that at nonzero
temperature the bound-states will be dissolved (dehadronized), but
the first necessary condition of the quark confinement criterion
will remain valid, nevertheless. In other words, quarks at nonzero
temperature, for example, in the quark-gluon plasma (QGP)
\cite{7}, will remain off-shell objects, i.e., by increasing temperature
they cannot be put on the mass-shell. Hence even in this case
they cannot be detected as physical particles (like electrons) in
the asymptotic states. That is why it is better to speak about
dehadronization phase transition in QGP rather than about
deconfinement phase transition.

Let us make a few remarks in advance. The region $c \ge x$ can be
considered as NP, whereas the region $c \le x$ can be considered
as the PT one (see next section). As was mentioned above, our
solutions to the quark propagator are valid in the whole momentum
range $[ 0, \infty)$. However, in order to calculate any physical
observable from first principles (represented by the corresponding
correlation function which can be expressed in terms of the quark
propagator integrated out), it is necessary to restrict ourselves
to the integration over the NP region $x \le c$ ($x \le c_0$)
only. This guarantees us that the above-mentioned unphysical
singularity (branch point at $x=c$ ($x = c_0$)) will not affect
the numerical values of the physical quantities. Evidently, this
is equivalent to the subtraction of the contribution in the
integration over the PT region $x \ge c$ ($x \ge c_0$). Let us
underline that at the hadronic level this is the necessary
subtraction which should be only made (see discussion in the next
section).

\section{Dynamical breakdown of chiral symmetry (DBCS)}

From a coupled system of the differential equations (2.7) it is
easy to see that this system

\vspace{3mm}

$allows \ a  \ chiral \ symmetry \ breaking \ solution \ only$,

\begin{equation}
\bar m_0 = 0, \quad A(x) \ne 0, \ B(x) \ne 0
\end{equation}
$and \ forbids \ a \ chiral \ symmetry \ preserving \ solution$,

\begin{equation}
\bar m_0 = B(x) = 0, \quad A(x) \ne 0.
\end{equation}
Thus any nontrivial solutions automatically break the $\gamma_5$
invariance of the quark propagator

\begin{equation}
\{ \gamma_5, S(p) \} = i\gamma_5 2 B(p^2) \neq 0,
\end{equation}
and they therefore $certainly$ lead to the  spontaneous chiral
symmetry breakdown at the fundamental quark level ($\bar m_0 = 0,
B(x) \ne 0$, the dynamical quark mass generation). Let us
emphasize that a measure of this breakdown is the twice
dynamically generated quark mass, i.e., at a scale of twice
dynamically generated quark mass a spontaneous chiral symmetry
breakdown occurs. In principle, one can calculate the dynamical
quark mass $B(p^2)$ at any finite point (at zero for regular
functions as in our case) and at any covariant gauge. However, the
important observation is that the above-mentioned definition of
this measure is valid for any covariant gauge, and in this sense
is gauge-invariant as it follows from the relation (4.3). In all
previous investigations a chiral symmetry preserving solution
(4.2) always exists. Here we do not distinguish between $B(p^2)$
and $\overline{B}(p^2)$ calling both dynamically generated quark
mass functions, for simplicity.

A few preliminary remarks are in order. A nonzero dynamically
generated quark mass function defined by conditions (4.1) and
(4.3) is the order parameter of DBCS at the fundamental quark
level. At the phenomenological level the order parameter of DBCS
is the nonzero chiral quark condensate defined as the integral of
the trace of the quark propagator in the chiral limit, i.e.,
(Euclidean signature, see Eq. (2.2))

\begin{equation}
< \bar q q>_0 = <0|\bar q q|0>_0 \sim i \int d^4p \ Tr S(p),
\end{equation}
up to unimportant (for our discussion) numerical factors. In terms
of the dimensionless variables (2.6) it becomes

\begin{equation}
< \bar q q>_0  \sim - \Lambda^3_{NP} \int_0^{\infty} dx \ x \
B_0(x),
\end{equation}
where for light quarks in the chiral limit $< \bar q q>_0 =  <
\bar u u>_0 =  < \bar d d>_0 =  < \bar s s>_0$, by definition.

It is worth emphasizing now that the phenomenological order
parameter of DBCS - the chiral quark condensate - defined as the
dynamically generated quark mass function $B_0(x)$ integrated out
(4.5) might be in principle zero even when the mass function is
definitely nonzero. Thus a nonzero dynamically generated quark
mass is a much more appropriate condition of DBCS than the chiral
quark condensate. One can say that this is the first necessary
condition of DBCS, while a nonzero chiral quark condensate is only
the second sufficient one.

However, this  is not the whole story yet. The problem is that the
chiral quark condensate defined in Eq. (9.5) still contains the
contribution in the integration over the PT region, say, $[y_0,
\infty)$. In order to define correctly the chiral quark condensate
in the INP QCD this contribution should be subtracted, i.e.,

\begin{equation}
< \bar q q>_0 \Longrightarrow < \bar q q>_0 + \Lambda^3_{NP}
\int_{y_0}^{\infty} dx \ x \ B_0(x) = - \Lambda^3_{NP}
\int_0^{y_0} dx \ x \ B_0(x).
\end{equation}
If now the mass function $B_0(x)$ is really the NP solution of the
corresponding quark SD equation, then this definition gives the
quark condensate beyond the PT theory. In our case this is so,
indeed. Moreover, it is easy to understand that in order to
guarantee that the algebraic branch point at $x = c_0$ will not
affect the numerical value of the quark condensate, the
dimensionless scale $y_0$, separating the NP region from the PT
one, should be identified with the constant of integration $c_0$.
Thus in our case the truly NP chiral quark condensate becomes

\begin{equation}
< \bar q q>_0 \sim - \Lambda^3_{NP} \int_0^{c_0} dx \ x \ B_0(c_0;
x),
\end{equation}
i.e., the truly NP dynamically generated quark mass function is
integrated out over the NP region as well. So there is not even a
bit of the PT information in this definition (all types of the PT
contributions have been already subtracted in Eq. (4.7)). The
point of subtraction at the hadronic level is determined by the
constant of the integration at the quark level (branch point).
Moreover, it depends on the mass gap $\Lambda_{NP}$ and not on the
arbitrary mass scales $1 \ GeV$, $2 \ GeV$, etc. In the PT limit
$\Lambda_{NP}^2 \rightarrow 0$ the quark condensate goes to zero
as it should be, by definition ($B_0(c_0; x)$ identically vanishes
in this case as well since, let us remind, $c_0, x \rightarrow
\infty$ uniformly). Thus in our approach the chiral quark
condensate itself has a physical meaning.

The actual calculation of the chiral quark condensate as well as
all other chiral QCD parameters will be given in the subsequent
paper, where the Goldstone sector of QCD will be analytically
investigated and numerically evaluated (for preliminary, however,
numerical results see our papers \cite{8,9}).

\section{Nonzero current quark masses. Light quarks}

To derive the explicit solutions for the quark propagator in the
general (non-chiral) case, it is convenient to start from the
ground system (2.7) itself, and rewrite it as follows:

\begin{eqnarray}
 x A' + (2 + x) A + 1 &=& - \bar m_0 B, \nonumber\\
2B B' + 3 A^2 + 2 B^2 &=& 2 \bar m_0 A B,
\end{eqnarray}
 where, let us remind, $A \equiv A(x)$, $B \equiv B(x)$, and
here the prime denotes the derivative with respect to the
Euclidean dimensionless momentum variable $x$. The dimensionless
current quark mass $\bar m_0$ is defined as $ \bar m_0 = m_0 /
\Lambda_{NP}$. We are interested in the solutions which are
regular at zero and asymptotically approach the free quark case at
infinity. Because of our parametrization of the quark propagator
(2.2), its asymptotic behavior has to be determined as follows
(Euclidean metrics):

\begin{equation}
A (x) \sim_{x \rightarrow \infty} - { 1 \over x}, \qquad B (x)
\sim_{x \rightarrow \infty} - { \bar m_0 \over x }.
\end{equation}
The ground system (5.1) is very suitable for numerical
calculations.

Let us now develop an analytical formalism which makes it possible
to find solution of the ground system (5.1) step by step in powers
of the light current quark masses, the so-called chiral
perturbation theory at the fundamental quark level. For this
purpose let us present the quark propagator form factors $A$ and
$B$ as follows:

\begin{eqnarray}
A (x) &=& \sum_{n=0}^{\infty} \bar m_0^n A_n (x), \nonumber\\
B (x) &=& \sum_{n=0}^{\infty} \bar m_0^n B_n (x),
\end{eqnarray}
and for light current quark masses one has $\bar m_0^{(u,d,s)} \ll
1$ (let us note in advance that this estimate is justified,
indeed). Substituting these expansions into the ground system
(5.1) and omitting some tedious algebra, one obtains

\begin{eqnarray}
 x A'_0(x) + (2 + x) A_0(x) + 1 &=& 0 , \nonumber\\
2B_0(x) B'_0(x)+ 3 A^2_0(x) + 2 B^2_0(x) &=& 0,
\end{eqnarray}
and for $n=1,2,3,...$, one has

\begin{eqnarray}
x A'_n(x) + (2 + x) A_n(x) &=& - B_{n-1}(x) , \nonumber\\
2P_n(x) + 3 M_n(x) + 2 Q_n(x) &=& 2 N_{n-1}(x),
\end{eqnarray}
where

\begin{eqnarray}
P_n (x) &=& \sum_{m=0}^n B_{n-m} (x) B'_m (x), \nonumber\\
M_n (x) &=& \sum_{m=0}^n A_{n-m} (x) A_m (x), \nonumber\\
Q_n (x) &=& \sum_{m=0}^n B_{n-m} (x) B_m (x), \nonumber\\
N_n (x) &=& \sum_{m=0}^n A_{n-m} (x) B_m (x).
\end{eqnarray}
Is is obvious that the system (5.4) describes the ground system
(5.1) in the chiral limit ($\bar m_0=0$), and thus coincides with
the system (2.12). Its exact solution is given by Eqs. (2.13) and
(2.14). The first nontrivial correction in powers of small $\bar
m_0$ is determined by the following system

\begin{eqnarray}
x A'_1 + (2 + x) A_1 &=& - B_0 , \nonumber\\
(B_1B'_0 + B_0 B'_1) + 3 A_0 A_1  + 2 B_0 B_1 &=& A_0 B_0,
\end{eqnarray}
where we omit the dependence on the argument $x$, for simplicity.
In a similar way can be found the system of equations to
determine terms of order $\bar m_0^2$ in the solution for the
quark propagator and so on.

Let us present a general solution of the first of Eqs. (5.5) as

\begin{equation}
 A_n(x) = - x^{-2} e^{-x} \int_0^x dx' \ x' e^{x'} B_{n-1} (x'),
\end{equation}
which is always regular at zero, since all $B_n(x)$ are regular as
well. The advantage of the developed chiral perturbation theory at
the fundamental quark level is that each correction in the powers
of small current quark masses is determined by the corresponding
system of equations which can be formally solved exactly.

Let us write down the system of solutions approximating the light
quark propagator up to first corrections, i.e.,

\begin{eqnarray}
A(x) &=& A_0(x) + \bar m_0 A_1(x) + ...., \nonumber\\
B(x) &=& B_0(x) + \bar m_0 B_1(x) + ....
\end{eqnarray}
This system is

\begin{equation}
A_0 (x) = x^{-2} (1 -x - e^{-x}), \quad A_0(0) = - {1 \over 2},
\end{equation}

\begin{equation}
 B^2_0(x) = 3 e^{-2x} \int_x^{c_0} dx' \  e^{2x'} A_0^2 (x').
\end{equation}
And

\begin{equation}
 A_1(x) = - x^{-2} e^{-x} \int_0^x dx' \ x' e^{x'} B_0 (x'),
\end{equation}

\begin{equation}
 B_1(x) = e^{-2x}B_0^{-1}(x) \int_{c_1}^x dz \ e^{2z} A_0(z)[ B_0(z)
 -3A_1(z)].
\end{equation}
In physical applications we also need $B^2(x)$, so we have

\begin{eqnarray}
B^2(x) &=& B_0^2(x) + 2 \bar m_0 B_0(x) B_1(x) + ..., \nonumber\\
&=& B_0^2(x) + 2 \bar m_0 e^{-2x} \int_{c_1}^x dz \ e^{2z} A_0(z)
[B_0(z) - 3A_1(z)] + ...,
\end{eqnarray}
and the relation between constants of integration $c_0$ and $c_1$
remains, in general, arbitrary. However, there exists a general
restriction, namely $B^2(x) \geq 0$ and real, which may lead to
some bounds for the constants of integration, however, that is $x
\leq c_0$ always remains valid.

\section{Nonzero current quark masses. Heavy quarks}

For heavy quarks it makes sense to replace $m_0 \rightarrow m_Q$,
i.e., to put $\bar m_Q = m_Q / \Lambda_{NP}$, and rewrite thus the
system of equations (5.1) as follows:

\begin{eqnarray}
x A' + (2 + x) A + 1 &=& - \bar m_Q B, \nonumber\\
2B B' + 3 A^2 + 2 B^2 &=& 2 \bar m_Q A B,
\end{eqnarray}
where, let us remind, $A \equiv A(x)$, $B \equiv B(x)$, and here
the prime denotes the derivative with respect to the Euclidean
dimensionless momentum variable $x$. We are again interested in
the solutions which are regular at zero and asymptotically
approach the free quark case at infinity (5.2), on account of the
above-mentioned replacement $m_0 \rightarrow m_Q$.

In this case it is convenient to find a solution for heavy quark
form factors $A$ and $B$ in the form of the corresponding
expansions, namely

\begin{eqnarray}
\bar m_Q^2A (x) &=& \sum_{n=0}^{\infty} \bar m_Q^{-n} A_n (x), \nonumber\\
\bar m_Q B (x) &=& \sum_{n=0}^{\infty} \bar m_Q^{-n} B_n (x),
\end{eqnarray}
and for heavy quark masses we have $\bar m_Q^{(c,b,t)} \gg 1$,
i.e., the inverse powers are small (let us note in advance that
this estimate is justified, indeed). In terms of the dimensionless
mass scale parameter $\bar m_Q$ the heavy quarks large mass limit
can be formally achieved by the two ways: $m_Q \rightarrow \infty$
at fixed $\Lambda_{NP}$ and at $m_Q$ fixed, while $\Lambda_{NP}
\rightarrow 0$.

Substituting these expansions into the first equation of the
ground system (6.1) and omitting some tedious algebra, one obtains

\begin{equation}
B_0(x) = -1, \quad B_1(x) = 0,
\end{equation}
and

\begin{equation}
xA'_n(x) + (2 + x)A_n (x) = - B_{n+2}(x), \quad n= 0,1,2,3,...
\end{equation}
In the same way, by equating terms at equal powers in the inverse
of heavy quark masses, from second of the equations of the ground
system (6.1), one obtains

\begin{eqnarray}
P_0(x) + Q_0(x) - N_0(x) &=& 0, \nonumber\\
P_1(x) + Q_1(x) - N_1(x) &=& 0.
\end{eqnarray}
and

\begin{equation}
P_{n+2}(x) + Q_{n+2}(x) - N_{n+2}(x) = -{3 \over 2}M_n(x),
  \quad n=0,1,2,3,...,
\end{equation}
where $P_n(z), \ M_n(z), \ Q_n(z), \ N_n(z)$ are again given
formally by Eqs. (5.6). Solving these equations, one obtains

\begin{eqnarray}
A_0(x) &=& B_0(x)= -1, \nonumber\\
A_1(x) &=& B_1(x) = 0,
\end{eqnarray}
and

\begin{eqnarray}
x A'_n(x) + (2+x) A_n(x) &=& - B_{n+2}(x), \nonumber\\
P_{n+2}(x) + Q_{n+2}(x) - N_{n+2}(x) &=& - {3 \over 2} M_n(x),
\quad n = 0,1,2,3, ...
\end{eqnarray}
It is possible to show that all odd terms are simply zero, i.e.,

\begin{equation}
A_{2n+1}(x)= B_{2n+1}(x) = 0,
 \quad n = 0,1,2,3, ...
\end{equation}
The explicit solutions for a few first nonzero terms are

\begin{equation}
A_0(x)= B_0(x) = - 1.
\end{equation}

\begin{eqnarray}
A_2(x) &=& x + {3 \over 2}, \nonumber\\
B_2(x) &=& x + 2.
\end{eqnarray}

\begin{eqnarray}
A_4(x) &=& - x^2 - {1 \over 2} x - 6, \nonumber\\
B_4(x) &=& - x^2 - {9 \over 2} x - 3.
\end{eqnarray}
Thus our solutions for the heavy quark form factors look like

\begin{equation}
A (x) = { 1 \over \bar m_Q^2} \sum_{n=0}^{\infty} \bar m_Q^{-n}
A_n (x) = - {1 \over \bar m_Q^2} + { x \over \bar m_Q^4} - {x^2
\over \bar m_Q^6} + ...+  D_A (x),
\end{equation}
where

\begin{equation}
D_A(x) = { 3 \over 2 \bar m_Q^4} - { x + 12 \over 2 \bar m_Q^6} +
...
\end{equation}
And

\begin{equation}
B(x) = { 1 \over \bar m_Q} \sum_{n=0}^{\infty} \bar m_Q^{-n} B_n
(x) = - {1 \over \bar m_Q} + { x \over \bar m_Q^3} - {x^2 \over
\bar m_Q^5} + ...+  D_B (x),
\end{equation}
where

\begin{equation}
D_B(x) = { 2 \over  \bar m_Q^3} - {9 x + 6 \over 2 \bar m_Q^5} +
...
\end{equation}
Summing up, one obtains

\begin{eqnarray}
A (x) &=& - { 1 \over x + \bar m_Q^2} + D_A (x), \nonumber\\
B(x) &=& - {\bar m_Q \over x+ \bar m_Q^2} +  D_B (x).
\end{eqnarray}

In terms of the Euclidean dimensionless variables (2.6), the quark
propagator (2.2) is

\begin{equation}
iS(x) = \hat x A(x) - B(x).
\end{equation}
Using our solutions, obtained above, it can be written down as
follows:

\begin{equation}
iS_h(x) = iS_0(x) + \hat x D_A(x)  - D_B(x),
\end{equation}
where $iS_0(x)$ is nothing but the free quark propagator with the
substitution $m_0 \rightarrow m_Q$, i.e.,

\begin{equation}
iS_0(x) = - { \hat x - \bar m_Q \over x + \bar m_Q^2}.
\end{equation}
Since $\hat x D_A(x)- D_B(x)$ is of the order $\bar m_Q^{-3}$,
then the heavy quark propagator (6.19) becomes

\begin{equation}
iS_h(x) = iS_0(x) + O(1 / \bar m_Q^3),
\end{equation}
which means that our solution for the heavy quark propagator is
reduced to the free quark propagator up to the terms of the order
$1 / \bar m_Q^3$.

\subsection{Heavy quark-gluon vertex}

For further purpose, it is instructive to go back to the
dimensional momentum variable $p$ in the heavy quark propagator,
i.e., to rewrite Eq. (6.21) up to the terms of the order $1 /
m_Q^3$ as follows:

\begin{equation}
S_h(p) = i { \hat p - m_Q \over p^2 + m_Q^2}.
\end{equation}
It is easy to derive that quantities defined in Eqs. (2.4) become

\begin{eqnarray}
\overline A(p^2) &=& -1, \nonumber\\
\overline B(p^2) &=& - m_Q, \nonumber\\
E(p^2) &=& (p^2  + m_Q^2)^{-1},
\end{eqnarray}
so that for the corresponding form factors (2.5) one gets

\begin{eqnarray}
 F_1(p^2) &=&  {1 \over 2}, \nonumber \\
 F_2(p^2) &=& F_4(p^2) = { 1 \over 2 m_Q}, \nonumber \\
 F_3(p^2) &=& 0.
\end{eqnarray}
Thus the explicit expression for the corresponding heavy
quark-gluon vertex (2.1) becomes

\begin{equation}
\Gamma_\mu(p,0) = { 1 \over 2} \Bigl[ \gamma_\mu + { 1 \over m_Q}
p_\mu + {1 \over m_Q} \hat p \gamma_\mu \Bigr].
\end{equation}
It is worth noting that an overall numerical factor $1/2$ in this
vertex is due to the second term in the ST identity (1.1), i.e.,
it is due to the fact that our initial identity is the ST one and
not QED-type in which the first term is to be only presented.

Concluding, a few remarks are in order. Starting from the
expansion (6.2) for the $A(x)$ function and using exact Eqs.
(2.10) and (2.11), one obtains the same free quark propagator
solution (6.17) for it. In other words, the straightforward
solution of the initial system (6.1) coincides with the exact
solution for the $A(x)$ function, indeed. Unfortunately, things
are not so simple for the heavy quark mass function $B(x)$.
Substituting the free quark propagator solution (6.17) for the
$A(x)$ function into the exact integral (2.8), on account of the
exact relation (2.9), one obtains the expression which is by no
means the solution of the free quark propagator for the $B(x)$
function shown in Eq. (6.17). It can be only reduced to it in the
direct $\bar m_Q \rightarrow \infty$ limit. Thus such obtained
solution is a new one for the heavy quark mass function, and it is
left to be investigated elsewhere.

\section{Conclusions}

We have investigated a closed system of equations for the quark
propagator obtained on the basis of our approach to low-energy QCD
which we call INP QCD. One of its general features is the
subtractions of all types and at all levels the PT contributions
("contaminations") from QCD. The above-mentioned system of
equations consists of the quark SD equation itself, which is
complemented by the quark ST identity for the corresponding
quark-gluon vertex. This system is free of all types of the PT
contributions at the fundamental quark-gluon-ghost level. Moreover, it is manifestly
gauge-invariant, i.e., does not depend explicitly on the gauge-fixing parameter.
It depends explicitly only on the mass gap responsible for the truly NP dynamics in
the QCD ground state. For the dynamically generated quark mass function $B(p^2)$
it admits an exact formal solution in terms of the $A(p^2)$ function, see Eq. (2.8).
The $A(p^2)$ function itself should satisfy the nonlinear differential equation
(2.11). This system of equations can be solved exactly in the chiral limit.
In the case of nonzero light quark masses, we develop an analytical formalism,
the so-called chiral perturbation theory at the fundamental quark level, which allows
one to find a solution for the quark propagator in powers of the light quark masses.
We also develop an analytical formalism, which allows us to find solution
for the quark propagator in the inverse powers of heavy quark masses. For the first
time it has been theoretically justified the use of the free quark propagator
for heavy quarks. We have established that this is possible even up to terms of the
order $1/m_Q^3$. Thus our solution automatically
possesses the heavy quark spin-flavor symmetry. However, we would like
to make once more perfectly clear the two main properties of
our solution for the quark propagator.

\subsection{Quark confinement}

One of the most important observations is that a formal
exact solution (2.8) for the dynamically generated quark mass
function has a branch point at $x=c$, which completely excludes a
pole-type singularity for the quark propagator similar to
the singularity which has electron propagator in QED. As mentioned above,
the quark confinement criterion in QCD consists of the two conditions.

\vspace{3mm}

{\bf I. The first necessary condition should be formulated at the
fundamental quark-gluon level as the absence of a pole-type
singularity in the quark propagator (see Eq. (3.1) and discussion therein).

\vspace{3mm}

II. The second sufficient condition should be formulated at the
hadronic level as the existence of a discrete spectrum only (no
continuum in the spectrum) in the bound-state problems.}

\vspace{3mm}

Though our solution for the quark propagator is formally valid in the whole
energy/momentum range, the integration out over the quark degrees of freedom should
be obviously made up to the branch point $x=c$ in order to prevent the quark propagator
to be pure imaginary in the region $x>c$. This is in agreement with the first
necessary condition. In this connection let us remind that nonconfining electron
propagator always has an imaginary part. In fact, the existence of the branch point
explicitly shows where the PT contributions are to be exactly subtracted when the quarks
degrees of freedom should be integrated out. For the gluon degrees of freedom
integrated out we already know the exact point of the subtraction \cite{10,11}.

At the macroscopic, hadronic level the linear rising potential
interpretation of confinement becomes relevant for bound states
between heavy quarks only. In this case the full vertex can be
approximated by its point-like counterpart (see Eq. (6.25) up to the
terms of the order $1/m_Q$). Saturating further the Wilson loop by the
one-gluon exchange diagram with the dominant $(q^2)^{-2}$ behavior
for the gluon propagator \cite{1,2}, one precisely obtains the
Wilson criterion of quark confinement--area law \cite{12,13}, or
equivalently the linear rising potential between heavy quarks.
In this case confinement looks more like that of the Schwinger model
\cite{14} of two-dimensional QED and of the 't Hooft model \cite{6}
of two-dimensional axial gauge QCD, where non-Abelian degrees of freedom
have been eliminated by the choice of the gauge.

Thus, the Wilson criterion of quark confinement and consequently
its linear rising potential interpretation based on lattice gauge
theory is valid only for heavy quarks, and, in general, is
inadequate for the continuous theory \cite{15}, indeed. At the same time,
a quark confinement criterion formulated above within the
continuous theory is valid for any quarks (light or heavy, does
matter). To analyze quark confinement in terms of the analytic
properties of the quark propagator is much more relevant. This
is reflected in the above-formulated its general criterion.
Heavy quarks demonstrate some special
properties, such as spin-flavor symmetry, linear rising potential,
a possibility to use the free quark propagators for them, etc. All this
substantially simplifies the investigation of the bound-states consisting of them.

Concluding, let us note that confinement has been proven for two-dimensional
covariant gauge QCD (i.e., taking into account non-Abelian degrees of freedom)
in our papers \cite{16}.

\subsection{DBCS}

The second important observation is that the $\gamma_5$ invariance of the quark
propagator is broken, for sure within our approach, see Eq. (4.3). As emphasized above,
this means the dynamical quark mass generation, and thus dynamical (or equivalently
spontaneous) breakdown of chiral symmetry. For the existence of this phenomenon
the two conditions are also required.

\vspace{3mm}

{\bf I. At the fundamental quark level the first necessary
condition is to be formulated as the absence of a chiral symmetry
preserving solution for the quark propagator,
while a chiral symmetry breaking solution is only allowed (see
Eqs. (4.2) and (4.1)).

\vspace{3mm}

II. At the phenomenological level the second sufficient condition
is to be formulated as the existence of the nonzero chiral quark
condensate, Eq. (4.7).}

\vspace{3mm}

The only problem with the nonzero quark condensate is that it
should be correctly calculated as explained above, i.e.,
the subtraction of all types and at all levels of the PT contributions should be
correctly done. This also assumes that it should depend on the mass gap
responsible for the NP dynamics in the QCD ground state. One can assign
a physical meaning to it now (since it does not depend on the arbitrary
scale) as the quantity which measures the density of quark degrees of freedom
in the vacuum \cite{17}. Precisely in this way the chiral quark condensate
was calculated within our approach (for preliminary numerical results see
our papers \cite{8,9}).

\subsection{INP QCD}

Let us briefly formulate some general features of INP QCD as
a theory of QCD at low energies.

(a). INP QCD is defined by the subtraction from QCD of all types and at
all levels of the PT contributions ("contaminations").

(b). First necessary subtraction should be made at the fundamental gluon
propagator level. The confining gluon propagator has been obtained by
the direct iteration solution of the gluon SD equation in the presence of a mass gap.

(c). Free from ghost complications because of the above-mentioned subtractions
in the ghost and quark-gluon sectors, though the information on the quark
degrees of freedom containing in the ghost-quark scattering kernel has been
self-consistently taken into account.

(d). Implies quark confinement.

(e). Implies DBCS.

(f). Has a uniquely defined physical mass gap responsible for color confinement,
DBCS and all other NP phenomena.

(g). Effectively short-range theory despite gluons remain massless.

(h). The chiral limit physics (i.e., the Goldstone sector) can be exactly evaluated.

(i). The point of the subtraction of the PT contributions at the hadronic
(macroscopic) level is exactly known. This is the branch point $x=c/x=c_0$ in the
general/chiral solution for the dynamically generated quark mass function (2.8)/(2.14).
Thus in this theory the NP (soft momenta) region is exactly separated
from the PT (hard momenta) one.

It is worth discussing this feature of INP QCD in more detail. The existence
of the point at which the PT tails should be exactly subtracted makes it
possible to establish the space of the smooth test functions, consisting of the quark
propagator and the corresponding quark-gluon vertex. In this space our generalized
function (the confining gluon propagator) becomes a continuous linear functional.
It is a linear topological space, denoted as $K(c)$ (in the chiral limit as $K(c_0)$),
consisting of infinitely differentiable functions having compact support in the
region $x \leq c$, i.e., test functions which vanish outside the
interval $x \leq c$ (in the chiral limit $x \leq c_0$). It is well known that this
space can be identified with a complete countably normed space, which is at
the same time a linear metric space, and the topology defined by the metric is
equivalent to the original topology \cite{18}. Thus the above-mentioned subtraction
of all kind of the PT contributions at the hadronic level becomes not only physically
well justified but mathematically confirmed by the distribution theory \cite{18}
as well. On the other hand, if one defines the space of the smooth test functions
as above, which is necessary because of the existence of the branch point, then
the subtractions of the PT tails become inevitable, indeed.

Concluding, a few remarks are in order. As underlined above,
INP QCD makes it possible to calculate the physical obsevables/processes in
low-energy QCD from first principles and in a self-consistent way. For this purpose
what one should mainly do is to express the $S$ matrix element of any physical quantity
in terms of the corresponding loop integral(s) over the derived confining quark
propagator and the corresponding bound-state amplitudes. These loop integrals
are integrals over the finite volume in which only the pure NP excitations and fluctuations
of the virtual transversal gluon/ghost and quark fields are important.
Due to the character of the above-established space of the smooth test functions,
the fluctuations of the PT character, origin and magnitude have been totally
"washed out" from this volume, and it remains free of them for ever.
For preliminary numerical calculations see, for example, the above-mentioned
Refs. \cite{8,9}).

The necessity to integrate up to the branch point, i.e., over the finite range only,
points out the existence of something like the Gribov horizon \cite{19,20},
established in the functional space. Our "horizon" exists in the momentum space,
which is much simper than the functional space. On the other hand, it has a clear
physical meaning as separating the NP "world" from the PT one in the true QCD vacuum.
It is well known that to solve the problem of an ambiguity in the gauge-fixing of
non-Abelian gauge fields (the so-called Gribov ambiguity (uncertainty), which results
in Gribov copies and vice-versa), he proposed to integrate over the finite range
in the functional space of non-Abelian gauge fields, which consists in integrating
only over the fields for which the Fadeev-Popov determinant is positive, introducing
thus the above-mentioned
horizon. Apparently, its existence in different spaces (functional, where
the Gribov copies problem is explicitly present, momentum, where it is implicitly
present, etc.) is inevitable at the macroscopic level. Otherwise this problem will
plague the dynamics of any essential non-linear gauge systems at all levels.
Within our approach the solution of the Gribov ambiguity problem in the gauge-fixing
of non-Abelian gauge fields at the fundamental (microscopic) gluon propagator level
is discussed in Ref. \cite{1}.

After performing the ultraviolet (UV) renormalization program for the mass gap
\cite{1} and making all necessary subtractions of the PT tails, this theory is evidently
UV finite, i.e., free of the UV divergences. Also, after performing the infrared
(IR) renormalization program for the mass gap \cite{1}, this theory becomes IR
renormalizable, i.e., free of severe IR singularities, expressed in terms of the IR
regularization parameter \cite{1,2}.
This is true at least for the considered sectors of the theory \cite{2} in order to derive
the confining quark propagator, which is needed for the
above-mentioned calculations from first principles. The general IR multiplicative
renormalizability of this theory is a rather technical issue and is left to be
proven elsewhere. It is beyond the scope of the present investigation.

\begin{acknowledgments}

Support in part by HAS-JINR Scientific Collaboration Fund and
Hungarian OTKA-T043455 grant (P. Levai) is to be acknowledged. I
would like to thank J. Nyiri for useful remarks, constant support
and help.

\end{acknowledgments}

\appendix

\section{Heavy quarks spin-flavor symmetry}

In order to investigate the heavy quarks symmetries (\cite{21} and
references therein) it is convenient to consider the heavy quark
propagator (6.22) and the corresponding heavy quark-gluon vertex
(6.25) in Minkowski space, since the investigation of the
above-mentioned symmetries includes hadron degrees of freedom (see
below). In Minkowski space they become

\begin{equation}
S_h(p) = i { \hat p + m_Q \over p^2 - m_Q^2}
\end{equation}
and

\begin{equation}
\Gamma_\mu(p,0) = { 1 \over 2} \Bigl[ \gamma_\mu + { 1 \over m_Q}
p_\mu - {1 \over m_Q} \hat p \gamma_\mu \Bigr].
\end{equation}

Let us explicitly show here that our solution for the quark
propagator in this case possesses the heavy quark flavor symmetry,
indeed. We will show that the quark propagator to leading order in
the inverse powers of the heavy quark mass will not depend on it,
i.e., it is manifestly flavor independent to the leading order of
this expansion. For this purpose, a standard heavy quark momentum
decomposition should be used, namely

\begin{equation}
p_{\mu} = m_Q \upsilon_{\mu} + k_{\mu},
\end{equation}
where $\upsilon$ is the four-velocity with $\upsilon^2=1$. It
should be identified with the four-velocity of the hadron. The
"residual" momentum $k$ is of dynamical origin. Substituting this
decomposition into the Eq. (A1), and taking into account only to
leading order terms in the inverse powers of $m_Q$, one finally
obtains

\begin{equation}
S_h(\upsilon, k) = {i \over \upsilon \cdot k } P_+ + O({1 \over
m_Q}),
\end{equation}
which is exactly the heavy quark propagator \cite{21}. Thus our
propagator does not depend on $m_Q$ to leading order in the heavy
quark mass limit, $m_Q \rightarrow \infty$, i.e., in this limit it
possesses the heavy quark flavor symmetry, indeed. The operator
$P_+$ is  a positive-energy projection operator, which satisfies
the following relations

\begin{equation}
P_{\pm} = {1 \over 2} ( 1 \pm \hat \upsilon), \quad P_{\pm}^2 =
P_{\pm}, \quad P_{\pm}P_{\mp} = 0.
\end{equation}

At the same time, the heavy-quark-gluon vertex (A2) in this limit
does not depend explicitly on the "residual" momentum $k$, i.e.,

\begin{equation}
\Gamma_\mu(\upsilon,0) = { 1 \over 2} \Bigl[ \gamma_\mu +
\upsilon_\mu - \hat \upsilon \gamma_\mu  +  O_{\mu}(1 / m_Q)
\Bigr].
\end{equation}

Let us now present some useful relations, namely

\begin{equation}
P_+ \gamma_{\mu}P_+ =P_+ \upsilon_{\mu}P_+, \quad P_+ \hat
\upsilon \gamma_{\mu}P_+ = P_+ \upsilon_{\mu},
\end{equation}
which can be easily checked. Because of these relations, the
heavy-quark-gluon vertex effectively becomes

\begin{equation}
\Gamma_\mu(\upsilon, 0) = \upsilon_\mu - {1 \over 2} P_+
\upsilon_\mu + O_{\mu}(1 / m_Q),
\end{equation}
i.e., the interactions of the heavy quark with light degrees of
freedom does not depend on its spin as well in the heavy quark
large mass limit (the dependence on the $\gamma$-matrix
disappears). This is a direct manifestation of the heavy quark
spin symmetry up to terms of the order $1/m_Q$. The existence of
the second term in this vertex is due to fact that our ST identity
is not trivial one, i.e., it is not QED-type as mentioned above.
Otherwise the only first term would survive (contribute). Taking
further into account the relation

\begin{equation}
P_+ P_+ \upsilon_{\mu}P_+ = P_+ \upsilon_{\mu}P_+,
\end{equation}
which is valid because of the second of relations presented in Eq.
(A7), the coupling of a heavy quark to gluons (A6) can be
effectively simplified further to

\begin{equation}
\Gamma_\mu(\upsilon, 0) =  {1 \over 2} \upsilon_\mu + O_{\mu}(1 /
m_Q),
\end{equation}
and again the coefficient $1/2$ is a manifestation that our
initial ST identity was not of QED-type one.

 Concluding, let us note that the general solution (2.8)
does not demonstrate the principal difference in the analytical
structure of its solutions for light and heavy quarks. At the
fundamental quark level the heavy quark mass limit is not Lorentz
covariant, and therefore for numerical calculations we will use
rather Eq. (2.8) than Eq. (6.17). In order to analyze a possible
symmetries of the interaction of heavy quarks with light degrees
of freedom, we will go to the heavy quark large mass limit ($m_Q
\rightarrow \infty$) at the final stage only within our approach.

\end{document}